\documentclass[10pt,twocolumn,english,conference,a4paper]{IEEEtran}
\usepackage[T1]{fontenc}
\usepackage[active]{srcltx}
\usepackage{color}
\usepackage{array}
\usepackage{float}
\usepackage{multirow}
\usepackage{amsthm}
\usepackage{amsmath}
\usepackage{graphicx}
\usepackage{gensymb}
\usepackage [latin1]{inputenc}

\makeatletter

\providecommand{\tabularnewline}{\\}
\floatstyle{ruled}
\newfloat{algorithm}{tbp}{loa}
\providecommand{\algorithmname}{Algorithm}
\floatname{algorithm}{\protect\algorithmname}

  \theoremstyle{plain}

\ifCLASSOPTIONcompsoc
\usepackage[caption=false,font=normalsize,labelfont=sf,textfont=sf]{subfig}
\else
\usepackage[caption=false,font=footnotesize]{subfig}
\fi

\usepackage{cite}
\usepackage{bm}
\usepackage{algorithmic}
\usepackage{algorithm}
\usepackage{graphicx}
\IEEEoverridecommandlockouts

\makeatother

\usepackage{babel}
\providecommand{\lemmaname}{Lemma}

\begin{document}

\title{\textcolor{black}{Systematic Performance Evaluation Framework for LEO Mega-Constellation Satellite Networks }}

\author{\IEEEauthorblockN{Yu Wang, Chuili Kong, Xian Meng, Hejia Luo, Ke-Xin Li, and Jun Wang \\}\IEEEauthorblockA{\textsuperscript{}Huawei
Technologies Co., Ltd., Hangzhou, China.\\
Email: wangyu207@huawei.com }} \vspace{-1.0em}
\maketitle
\begin{abstract}
\textcolor{black}{Low Earth orbit (LEO) mega-constellation satellite networks have shown great potential to extend the coverage capability of conventional terrestrial networks. How to systematically define, quantify, and assess the technical performance of LEO mega-constellation satellite networks remains an open issue. In this paper, we propose a comprehensive key performance indicator (KPI) framework for mega-constellation based LEO satellite networks. An efficient LEO constellation oriented performance evaluation methodology is then carefully designed by resorting to the concept of interfering area and spherical geographic cell. We have carried out rigorous system-level simulations and provided numerical results to assess the KPI framework. It can be observed that the achieved area traffic capacity of the reference LEO constellation is around 4 Kbps/km$^2$, with service availability ranging from 0.36 to 0.39.  Besides, the average access success probability and handover failure rate is approximate to 96\% and 10\%, respectively, in the nearest satellite association scheme.}\end{abstract}
\begin{IEEEkeywords}
 LEO satellite constellation, KPI, performance evaluation, quasi-earth-fixed, beam hopping, 3GPP NTN. \textmd{\normalsize{\vspace{-0.5em}}}{\normalsize \par}
\end{IEEEkeywords}

\section{Introduction\label{sec:introduction}} \vspace{-0.5em}

\textcolor{black}{Next generation communication era is expected to assure three-dimensional global  wireless connectivity and bridge the digital divide through seamlessly integrating
non-terrestrial networks (NTNs) \cite{Survey_NTN5Gto6G,Survey_SatComintheNewSpaceEra,VLEO_HW}. To accomplish the ambitious vision, innovative standardization endeavors have been sponsored by
the Third Generation Partnership Project (3GPP) to study a set of necessary adaptations enabling the operation of 5G New Radio (NR)
protocol in the NTN context  \cite{38821_SolutionstoSupportNTN,38811}.  Specifically,  LEO mega-constellation satellite networks (LMCSNs) represent a burgeoning frontier scenario for NTN, and thus have recently attracted substantial academical and industrial interests. Composed by a large number of inter-connected LEO satellites (typically several thousands or even more), LMCSNs can provide increased network coverage, improved broadband capacity, and reduced end-to-end delay. Numerous commercial solutions have been envisaged to provide  broadband Internet access by deploying LEO satellite mega-constellations, e.g.,
Oneweb, Kuiper, Starlink and AST SpaceMobile.}

\textcolor{black}{Before delving into LMCSN, we should first address the fundamental issue of what critical capability the network can provide. As such, it is requisite to define and characterize the corresponding key performance indicators (KPIs) therein. On one hand, a well known KPI framework has been put forward in authoritative standard organizations such as International Telecommunication Union (ITU) \cite{ITUR_M2514} and 3GPP \cite{38913}. However, the baseline KPI framework can not fully reflect the distinct characteristic for LMCSNs in terms of access, mobility, and constellation-specific networking performance. On the other hand, there are some pioneering efforts  devoted to investigating LMSCN-specific KPIs. More specifically, satellite handover (HO) related KPIs (e.g., time-of-stay, and radio link failures) \cite{Nokia_HOSolutionsforLEO} are evaluated based on system-level simulation for earth-moving LEO satellite networks. In \cite{ConstellationDesign}, the LEO constellation design problem is studied to optimize constellation KPIs such as service availability and scalability. A new performance metric, i.e., service coverage, defined as the ratio of traffic density to service requirement per unit area is introduced in \cite{ChinaCom_6GServiceCoverage} to describe the coverage capability for LMSCNs. Besides, by resorting to stochastic geometry, the downlink coverage probability and average data rate are theoretically analyzed for LEO constellations. \textit{To sum up, state-of-the-art studies only concentrate on partial aspects of the KPI system, and a systematic  KPI framework for LEO mega-constellations is still missing in the literature.}}
\begin{figure}
\centering \vspace{-1.0em} \includegraphics[scale=0.5]{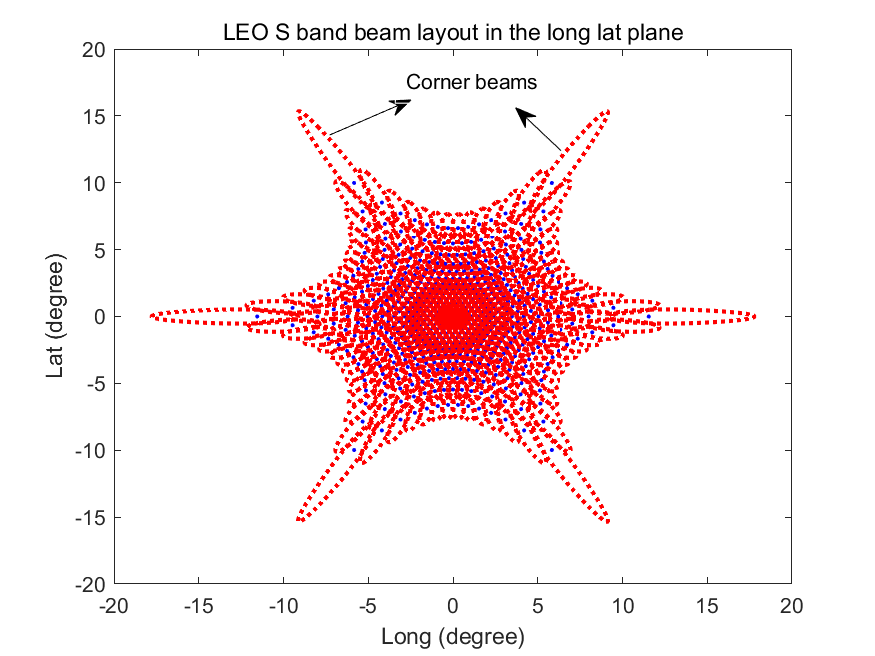}\vspace{-0.5em}
\caption{Beam layout in existing single-satellite simulation.\label{fig:NTNModeling}} \vspace{-2.0em}
\end{figure}

\textcolor{black}{Accordingly, to evaluate the aforementioned KPI system for the emerging LMCSN scenario, an efficient performance assessment methodology is necessary. Considering the distinct satellite network feature such as frequency/polarization reuse and beam layout configurations, an extension of conventional wrap-around mechanism with additional surrounding beams is applied in case of  single-satellite interference modeling \cite{38821_SolutionstoSupportNTN,ITUR_M2514}. However, the above approach is only effective in the specific case of central beams at satellite nadir (i.e., around 90\degree\,elevation angle), and cannot be directly extended to the multi-satellite simulation case. This is because the beam layout is defined by hexagonal mapping of the beam bore sight directions on UV plane, and the generated beams in the geodetic plane exhibit severe distortion especially at the edge (corner beams as depicted in Fig. \ref{fig:NTNModeling}) within satellite coverage area. To solve the problem, \cite{38821_SolutionstoSupportNTN} presents two alternative evaluation solutions for multi-satellite scenario. \textit{Unfortunately, both the realistic beam layout configuration details and multi-satellite interference modeling technique under highly dynamic network environment are not explicitly investigated. Therefore, more endeavors should be made for thorough constellation-wise performance assessment in LMCSNs.}}

\textcolor{black}{To circumvent the above issues, a systematic KPI framework including the target KPI set definition and relevant evaluation methodology should be devised for LEO satellite networks. However, it is technically challenging to develop a comprehensive KPI framework for LMCSNs, due to several reasons: 1) In addition to  radio interface technology (RIT) related KPIs, the impact on access and mobility related performance should be characterized in the specific satellite scenario. More importantly, constellation KPIs need to be exploited to evaluate the performance of a whole satellite constellation;
2) A LEO mega-constellation exhibits the following distinct characteristics such as no near-far effect and irregular beam layout, and thus the interference problem in satellite environment becomes critical. To make the matter worse, the fast mobility of LEO satellites induces the complicated and time-varying multi-satellite interference problem; 3) It is prohibitive to simulate a complete mega-constellation taking into account the large time-space span of the network. To strike a good balance between simulation complexity and modeling fidelity, an efficient performance evaluation scheme is necessary \cite{WCM_SAGIN_Simulation}.}

To fill in this gap, in this paper, we try to answer these challenging questions for LMCSNs. \textit{To the authors' best knowledge, this is the first work in the literature to define and assess the KPI system especially for LEO mega-constellation networks.} The main contributions are summarized as follows.
\begin{itemize}
  \item By referring to the ITU/3GPP KPI framework, we further propose a comprehensive KPI system to characterize the specific satellite network performance. The proposed KPI system can fully capture the constellation capability, access capacity, and mobility performance for LMCSNs.
  \item A highly-efficient constellation-wise performance evaluation methodology is devised through exploitation the concept of interfering area and hexagonal spherical cells (SCs), such that multi-satellite interference can be approximately modeled in the dynamic and complex LMCSN environment. The evaluation scheme can achieve a proper assessment of LMCSN with reasonable simulation complexity while not sacrificing too much modeling fidelity. This evaluation methodology will establish the base for further deep investigation of LMCSNs.
  \item We have developed a versatile system-level simulator that is rigorously calibrated upon many ITU/3GPP baseline configurations. Extensive experiments are implemented and numerous simulation results of target KPIs are present. Under practical LMCSN simulation configurations, some system design insights and quantitative performance limitations can be found. For instance, the achieved area traffic capacity of the reference LMCSN with 1800 satellites is only around 4 Kbps/km$^2$, with service availability ranging from 0.36 to 0.39.
\end{itemize}

The remainder of this paper is structured as follows. Section \ref{sec:KPI-Framework}
introduces the system model and presents the KPI framework in detail.
The LMCSN-specific evaluation methodology is elaborated in Section \ref{sec:Evaluation-methodology}. Section \ref{sec:Performance-Evaluation}
provides the simulation results, followed by conclusions in Section
\ref{sec:Conclusion}.

\section{KPI Framework Definition \label{sec:KPI-Framework}}

In this section, we describe the system model under consideration and the proposed KPI framework for LMCSNs. \vspace{-0.5em}
\subsection{System Model}

We consider a LMCSN operating at a center frequency $f_{0}$ with total system bandwidth $W$. The LMCSN comprises a set of $\mathcal{L}=\{1,\ldots,L\}$ LEO satellites. Each satellite $l$ is equipped with $M = M_xM_y$ phased array antennas to flexibly generate a maximum of $B_l$ beams, where $M_x$ and $M_y$ denote the number of antennas along the x- and y-axes, respectively. Without loss of generality, the antenna elements are equally spaced by 0.5 wavelength along both axes.

\begin{figure}
\centering \vspace{-1.0em} \includegraphics[scale=0.75]{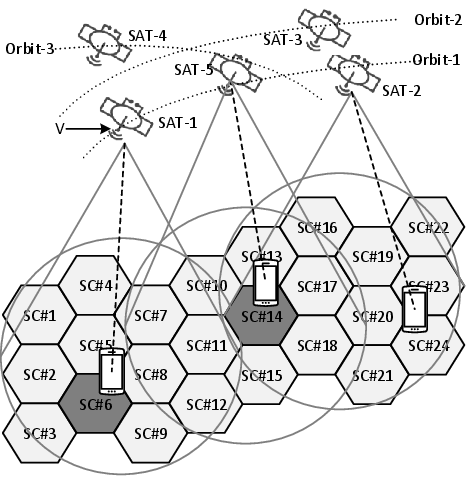}\vspace{-0.5em}

\caption{An example quasi-earth-fixed LEO NTN system.\label{fig:LEO-NTN-system} } \vspace{-1.5em}
\end{figure}

 As adopted in commercial LEO constellations, e.g., Starlink and AST, the satellite beams work in the quasi-earth-fixed mode. That is, beams continuously cover one geographic area for a limited period, and turn to cover a different geographic area during another period. The target coverage area is divided into  $\mathcal{C}=\{1,\ldots,C\}$  SCs, wherein a SC corresponds to a fixed hexagonal geographical area on the Earth's surface. In addition, SC-centric beam hopping method \cite{ICC_CooperativeBeamHopping,BH_Simulation} is applied to serve the wide coverage area. To be specific, once being activated, a beam points to the center position of a target SC. An example LMCSN system with 5 satellites and 24 SCs is shown in Fig. \ref{fig:LEO-NTN-system}, wherein only 2 SCs (i.e., SC\#6 and SC\#14) are illuminated by satellite beams at the time period.

A set of  $\mathcal{U}=\{1,\ldots,U\}$ UEs are distributed in the target area. Time horizon is discretized into $\mathcal{T}=\{1,\ldots,T\}$ equal time slots. We define a binary scheduling variable $\alpha_{b,l}^{c}(t)=1$  if beam $b$ of satellite $l$ is illuminated to SC $c$ at time slot $t$, and $\alpha_{b,l}^{c}(t)=0$ otherwise. The received power $P_{b,l}^{u}(t)$ for UE $u$ served by beam $b$ of satellite $l$ at time slot $t$ is derived as \vspace{-1.5em}

\begin{equation}
P_{b,l}^{u}(t)= P_T+G_T(\theta_{b,l}^u(t),\phi_{b,l}^u(t))+G_R-PL(d_{b,l}^u(t)),
\end{equation}
where $P_T$ is the beam transmit power, $G_R$ corresponds to the UE receive gain, and $G_T(\theta_{b,l}^u(t),\phi_{b,l}^u(t))$ represents the transmit antenna gain \cite{BH_Simulation}, with $\theta_{b,l}^u(t)$ and $\phi_{b,l}^u(t)$ denoting the elevation angle and azimuth angle, respectively. Besides, $PL(d_{b,l}^u(t))$ denotes the total path loss (in dB), which is further calculated as specified by 3GPP NTN specifications \cite{38811} \vspace{-1.5em}

\begin{equation}
PL(d_{b,l}^u(t))=32.45+20\log_{10}(f_{0}\cdot d_{b,l}^u(t))+F_s+L_{g}+L_{s},
\end{equation}
wherein  $d_{b,l}^u(t)$ is the slant path distance, $F_s$ represents the log-normal distributed shadow fading, and $L_{g}$ and $L_{s}$ represent the atmospheric absorption and scintillation
loss, respectively.

Consequently, the overall signal-to-interference plus noise ratio (SINR) $\gamma_{b,l}^{u,c}(t)$ of UE $u$ located in SC $c$ from serving beam $b$ of satellite $l$ at time slot $t$ is denoted by \vspace{-0.5em}

\begin{equation}
\gamma_{b,l}^{u,c}(t)=\frac{P_{b,l}^{u}(t)\alpha_{b,l}^{c}(t)}{I_{intra}+I_{inter}+N_{0}},\label{eq:SINR}
\end{equation}
where $I_{intra}=\sum_{b^{'}\neq b,l^{'}=l,c^{'}}P_{b^{'},l^{'}}^{u}(t)\alpha_{b^{'},l^{'}}^{c'}(t)$ is the intra-satellite interference, $I_{inter}=\sum_{b^{'},l^{'}\neq l}P_{b^{'},l^{'},c'}^{u}(t)\alpha_{b^{'},l^{'}}^{c'}(t)$ is the inter-satellite interference, and $N_{0}$ is the noise power determined by UE noise figure and antenna temperature.

\subsection{Proposed KPI System for LMCSN}
In order to identify the new services, capabilities, and minimum technical performance requirements, a set of representative and multifaceted KPI parameters should be accordingly exploited. In particular, the KPI system defined in ITU/3GPP \cite{ITUR_M2514,38913} comprises the following principal aspects such as peak data rate, user experienced data rate, latency, mobility, connection density, energy efficiency, spectrum efficiency, and area traffic capacity. As can be seen, the KPI parameters are envisaged only from RIT's perspective.

Motivated by this, the following LMCSN-specific KPIs are carefully selected, which can be further categorized into constellation KPIs and RIT KPIs as summarized in Table \ref{tab:KPI-systems}. To be specific, constellation KPIs are employed to reflect the performance of a satellite constellation.
\begin{itemize}
  \item  \textit{N-asset coverage}: the number of simultaneously serviceable satellites per geographic area wherein the received satellite signal quality is above a predefined threshold, e.g., the perceived SNR at a target SC from each of the $N$ satellites is better than a predetermined threshold. The KPI is introduced to characterize the multi-fold constellation coverage performance for a LMCSN.
  \item  \textit{Area traffic capacity}: the total traffic throughput served per geographic area. This is a measure of how much traffic a network can carry per unit area.
  \item  \textit{Service availability}: the probability that a (service) beam can be immediately scheduled to serve a target geographic area. It is worth noticing that the KPI is carefully selected to capture the LMCSN-specific beam hopping service ability, i.e., how soon a specific SC can be served again by at least a satellite beam.
\end{itemize}

\begin{table}[t]  \vspace{-1.0em}
\textcolor{black}{\caption{\textcolor{black}{Selected KPI set for LEO satellite networks\label{tab:KPI-systems}. }} \vspace{-1.0em}
}\centering\textcolor{black}{}%
\begin{tabular}{|l|l|l|}
  \hline
  \textbf{Category} & \textbf{Characteristic}            & \textbf{Evaluation} \\ \hline
                    & N-asset coverage                   & Simulation \\  \cline{2-3}
  Constellation KPI & Area traffic capacity              & Simulation \\ \cline{2-3}
                    & Service availability               & Simulation \\ \cline{1-3}
                    & Peak data rate                     & Analytical\\ \cline{2-3}
                    & User experienced data rate         & Simulation  \\ \cline{2-3}
                    & Unmet capacity                     & Simulation\\ \cline{2-3}
                    & Energy efficiency                  & Inspection\\ \cline{2-3}
   RIT KPI          & User plane latency                 & Analytical\\ \cline{2-3}
                    & Control plane latency              & Analytical\\ \cline{2-3}
                    & Access success probability         & Simulation\\ \cline{2-3}
                    & Access capacity                    & Simulation\\ \cline{2-3}
                    & Mobility interruption time         & Simulation\\\cline{2-3}
                    & Handover failure rate              & Simulation\\
  \hline
\end{tabular} \vspace{-1.5em}
\end{table}

Meanwhile, similar to the definition in conventional cellular networks, RIT KPIs are also adopted to reflect the performance, e.g., access capacity and mobility performance, of possibly multiple cells within each satellite.
\begin{itemize}
  \item  \textit{Peak data rate}: the highest theoretical data rate assuming error-free conditions and all assignable radio resources. We use this conventional KPI to quantify the satellite transmission capacity.
  \item \textit{User experienced data rate}: achievable data rate that is available ubiquitously across the coverage area to a mobile UE. We utilize this conventional KPI to reflect the UE perceived data rate experience in practice.
  \item \textit{Unmet capacity}: the gap between required UE data rate and offered data rate. The KPI is chosen herein to demonstrate the effect of unbalanced spatial-temporal traffic distribution on network service satisfaction degree.
  \item \textit{Energy efficiency}: the quantity of information bits per unit of energy consumption of the radio access network/device in bit/Joule. This is a sustainability KPI.
  \item \textit{User plane latency}: the time it takes to successfully deliver an application layer packet/message from the ingress point to the egress point. The KPI quantifies how fast a data transmission can be achieved. Notably, propagation delay should be accounted in computing the user plane latency for LMCSNs.
  \item \textit{Control plane latency}: This refers to the time to move from a battery efficient state to start of continuous data transfer. Similarly, propagation delay is a non-negligible factor when analyzing the control plane latency.
  \item \textit{Access success probability}: the probability to successfully complete the random access procedure within the maximum number of preamble transmissions. We capitalize this KPI to measure the RACH relevant capability.
  \item \textit{Access capacity}: the number of successfully accessed UE number per geographical area. In association with access success probability, access capacity is proposed to quantify the critical LMCSN  access capability.
  \item \textit{Mobility interruption time}: the shortest time duration supported by the system during which a UE cannot exchange user plane packets with any satellite during HO transitions. This is a mobility related measure.
  \item \textit{Handover failure rate}: If a UE does not HO to another cell despite the weak signal quality of the serving cell, the UE experiences a HO failure. The HO failure rate is defined as the ratio of HO failure number to total HO number. We employ the KPI to show the performance of high network mobility triggered frequent HOs.
\end{itemize}  \vspace{-0.5em}

\section{Evaluation Methodology \label{sec:Evaluation-methodology}}

In this section, an overview of the proposed evaluation methodology is firstly given. After that, the key modules underlying the evaluation scheme are discussed. \vspace{-0.5em}

\subsection{Overview of the Proposed Evaluation Scheme}

\begin{figure}[t]
\centering \vspace{-1.0em} \includegraphics[scale=0.95]{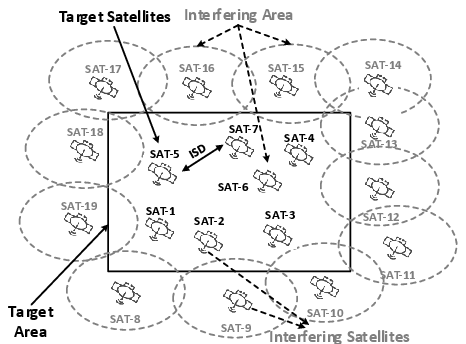}\vspace{-0.5em}

\caption{\ LEO constellation evaluation methodology overview. \label{fig:InterferingArea} } \vspace{-2.0em}
\end{figure}

\textcolor{black}{Generally, it is complex and time-consuming to simulate an entire mega-constellation. To achieve a tradeoff between simulation complexity and modeling fidelity, a highly efficient evaluation scheme for the KPI system is indispensable. In light of this, we exploit the concepts of interfering area and SCs for assessing the multi-satellite case. The outline of the proposed evaluation methodology is described as follows. \\
\textbf{Step 1):} Define a reference LEO constellation (walker polar constellation and/or walker delta constellation) with detailed constellation parameters such as orbit height, orbit inclination, number of orbits, and number of satellites per orbit. Note that full constellation simulation is not required, at least for the initial calibration process in the system-level simulation. \\
\textbf{Step 2):} Set one or more target areas (e.g., low-latitude area and/or high-latitude area) for simulation. The set of SCs can be obtained for the target area based on the H3 hexagonal hierarchical geospatial indexing system \cite{H3}. The cell radius of the SC is determined by the beam radius at satellite nadir. \\
\textbf{Step 3):} Divide the simulated time duration into $S$ snapshots with equal time length. The network topology, e.g., the inter-satellite distance (ISD), is fixed in each snapshot, and dynamically updated during snapshot transitions.  The snapshot duration (e.g., 10s) is typically much larger than that of a time slot (e.g., 1ms) as defined in Section \ref{sec:KPI-Framework}-A. The two time scale configuration can reduce the simulation complexity. \\
\textbf{Step 4):} Select a subset of $N$ target satellites, and determine the associated interfering area for approximate interference analysis. The rule for choosing the interfering area will be elaborated later in Section \ref{sec:Evaluation-methodology}-B. Both the target satellite set and interfering area are renewed as snapshot iterates.   \\
\textbf{Step 5):} Collect the KPI statistics only for the example $N$ target satellites. KPI statistics from other satellites in the interfering area are not required. To this end, the computational load of the system-level simulations is further decreased.} \vspace{-0.5em}

\subsection{Key Module Analysis}
Herein, we investigate the key modules for assessing the proposed KPI system. Compared with traditional cellular networks, the following key modules should be paid special attention and explicitly modeled in the LMCSN simulation.

\begin{itemize}
  \item  \textit{Module I: Multiple-satellite interference modeling}: To model the multi-satellite interference, interfering area is carefully defined. All satellites (including the serving satellite itself) located inside the interfering area should be accounted for interference calculation. The interference is calculated using dynamic ISD based network topology and realistic beam layout generated by the phased array antenna pattern. It should be emphasized that besides the target area, the interfering area should at least contain the area covered by a complete surrounding layer of interfering satellites outside the target area.  As shown in Fig. \ref{fig:InterferingArea}, the interfering area comprises two components. One is the target area that consists of 7 target satellites, i.e., \{SAT-1,...,SAT-7\}. The other is the surrounding area with 12 interfering satellites, i.e., \{SAT-8,...,SAT-19\}. Interference from all the 19 satellites is computed for any target satellite inside the target area.
  \item  \textit{Module II: Dynamic topology modeling}: Snapshot based topology generation is adopted to characterize the dynamic evolution of network connectivity. With the change of snapshots, the satellite locations and ISDs are updated using popular orbit propagation models, e.g., Keplerian model, and Secular J2 model. 
  \item  \textit{Module III: Large propagation delay modeling}: The propagation delay is a non-negligible factor in NTN scenario. It is computed based on real-time UE location and satellite ephemeris in each time snapshot. Besides, enhancing features such as timing relationship and HARQ process should be designed for the simulation.
  \item  \textit{Module IV: Unbalanced traffic modeling}: The unbalanced and area-specific traffic distribution is common in satellite scenario. To model this effect, we generate non-uniform service requests by defining hotspot and non-hotspot areas, where different levels of UE density are used.
\end{itemize}

\section{Performance Evaluation\label{sec:Performance-Evaluation}}

In this section, practical simulation settings following 3GPP NTN assumptions
are configured. Afterwards, numerical results are presented
to evaluate user/control plane related KPIs. \vspace{-1.5em}

\subsection{Simulation Settings}

\begin{table}[t] \vspace{-1.5em}
\textcolor{black}{\caption{Key simulation parameters\label{tab:Key-simulation-paramters.}. }} \vspace{-1.0em}
\centering\textcolor{black}{}%
\begin{tabular}{|l|l|}
\hline
\textbf{Parameters}  & \textbf{Values}\tabularnewline
\hline
\textcolor{black}{The number of orbit }  & \textcolor{black}{60}\tabularnewline
\hline
The number of satellite per orbit   & 30\tabularnewline
\hline
Orbit inclination  & \textcolor{black}{55}\tabularnewline
\hline
\textcolor{black}{Orbit height  } & {508} km\tabularnewline
\hline
Available broadcasting beams     & 5 \tabularnewline
\hline
Available service beams    & 50 \tabularnewline
\hline
Broadcast beam EIRP density    & 32.29 dBW/MHz \tabularnewline
\hline
Service beam EIRP density   & 41.41 dBW/MHz  \tabularnewline
\hline
Antenna configuration for broadcasting beams     & 7$\times$7 \tabularnewline
\hline
Antenna configuration for service beams    & 20$\times$20 \tabularnewline
\hline
Carrier frequency  & 3.65 GHz\tabularnewline
\hline
System bandwidth  & 30 MHz\tabularnewline
\hline
Channel model & Clear sky with LOS\tabularnewline
\hline
Scintillation loss and other loss & 5.5 dB\tabularnewline
\hline
UE antenna G/T  & -33.62 dB/K\tabularnewline
\hline
Packet size & 0.5 MByte\tabularnewline
\hline
UE attachment & RSRP\tabularnewline
\hline
Radio resource scheduler & RR+PF scheduler\tabularnewline
\hline
PRACH occasion periodicity & 80ms \tabularnewline
\hline
The number of preambles for HO/initial access & 10/54 \tabularnewline
\hline
\end{tabular} \vspace{-2.0em}
\end{table}

We use a self-developed system-level simulator with C++ programming for the technical performance evaluation. The simulator has been accurately calibrated under massive baseline configurations defined by 3GPP/ITU. With the simulator, we have provided the first and complete throughput performance results among all participant companies in the 3GPP NTN Release 16 study phase. The throughput results are captured in the formal technical report \cite{38821_SolutionstoSupportNTN}.

In the simulation, a LMCSN consisting of 1800 satellites is considered. The target area with longitude and latitude setting to [90,110] and [25,45], respectively, is selected for simulation. Satellites whose nadir points located inside the target area are chosen as target satellites, while the interfering area corresponds to the area covered by all satellites visible to the target area. The H3 geospatial indexing system \cite{H3} is employed to create a set of SCs, wherein the cell radius corresponding to broadcasting beams and service beams is 59.8km and 22.6km, respectively. Within the target area, 10 SCs are randomly selected as hotspot SCs, while the other SCs are non-hotspot SCs. For traffic generation, different number of UEs are deployed depending on the SC type. The number of UEs per hotspot SC is set to 500, and the number of UEs per non-hotspot SC is chosen from the set \{100,300\} depending on the scenario configuration. Each UE generates session requests following poisson arrival rate of 1/300 and average session duration of 30 seconds.  The time span for simulation is 6000 seconds, which is further divided into 600 equal snapshots. The set of key parameters are summarized in Table \ref{tab:Key-simulation-paramters.}.

For user plan capability related simulations, the round-robin (RR) beam hopping scheme with interference avoidance is adopted as the baseline \cite{BH_Simulation}. As in \cite{38821_SolutionstoSupportNTN}, the classical  proportional fairness (PF) scheduler with 1ms time slot length is utilized for radio resource management. While for simulating control plane related capabilities such as access and mobility, two benchmark schemes are adopted. The first scheme is termed as Nearest scheme, where each SC is only served by its nearest satellite. UEs located in a SC are required to perform HO at the next snapshot if their serving satellite turns to change. The second scheme is termed as SSB-plan-nearest scheme, where each SC is covered by two nearest satellites to ensure seamless HO. The triggering event for HO is based upon both time CondEvent T1 and signal quality Event A4 as defined in \cite{TS38331}. In CondEvent T1, the HO time duration is determined as the snapshot when the source satellite cannot provide broadcasting beams (i.e., SSBs) for current SCs. In signal quality Event A4, the HO can be executed when the neighbor satellite's SSB-SINR becomes better than -6dB. To further reduce interference, the two satellites coordinate to illuminate broadcasting beams for a given SC at different time slots as much as possible. \vspace{-0.5em}

\subsection{System-Level Simulation Results}
\begin{figure}[t]
\centering \vspace{-1.0em} \includegraphics[scale=0.4]{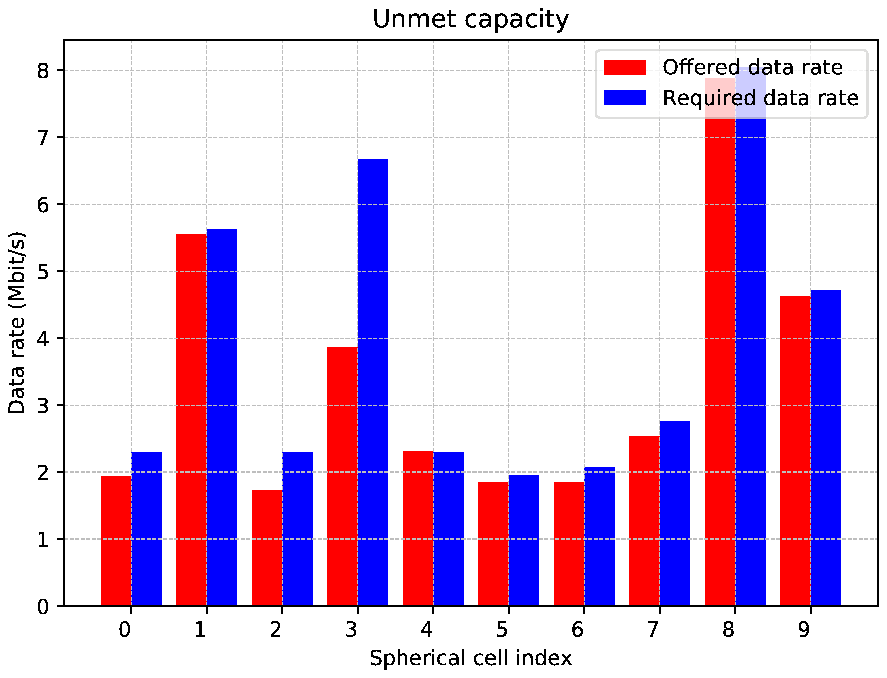} \vspace{-0.5em}
\caption{Unmet capacity performance of 10 sampling SCs.\label{fig:unmet_capacity} } \vspace{-1.0em}
\end{figure}

\begin{table}[t]
\textcolor{black}{\caption{\textcolor{black}{Constellation specific KPI simulation results\label{tab:constellation-KPI}. }} \vspace{-1.0em}
}\centering\textcolor{black}{}%
\begin{tabular}{|l|l|l|l|}
\hline
\textbf{Characteristic}  & \textbf{5\%} & \textbf{50\%} & \textbf{95\%}\tabularnewline
\hline
\textcolor{black}{$N$-asset coverage (1)} & 7 & 9 & 11  \tabularnewline
\hline
Area traffic capacity (Kbps/$km^2$)& 3.96 & 4.34  & 4.48  \tabularnewline
\hline
Service availability (\%)& 0.36 & 0.37 & 0.39 \tabularnewline
\hline
\end{tabular} \vspace{-2.0em}
\end{table}

\begin{figure}[t]
\centering \vspace{-1.0em} \includegraphics[scale=0.45]{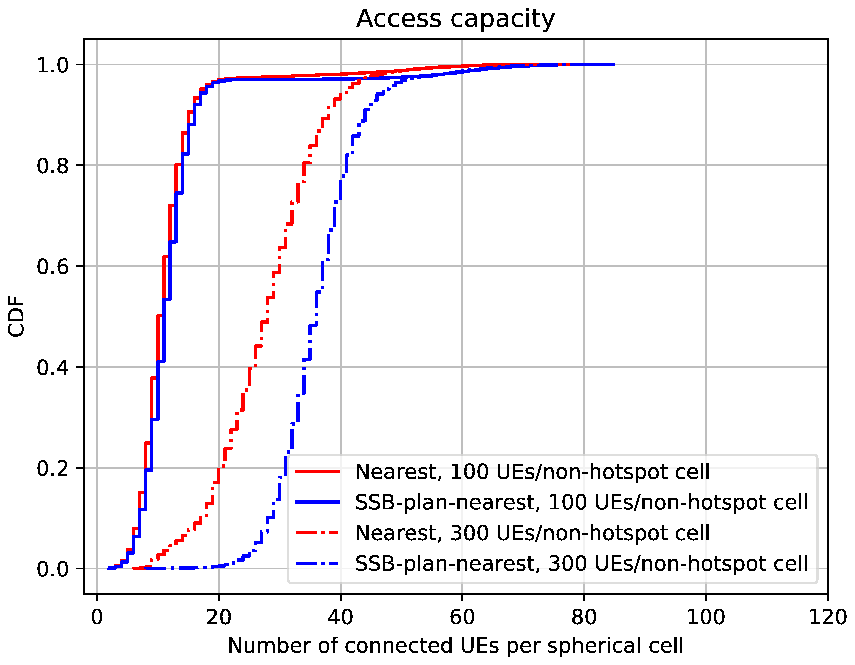} \vspace{-1.0em}
\caption{CDF of access capacity in target SCs.\label{fig:access_capacity} } \vspace{-1.0em}
\end{figure}

\textcolor{black}{}
\begin{figure}[htp]
\centering \includegraphics[scale=0.32]{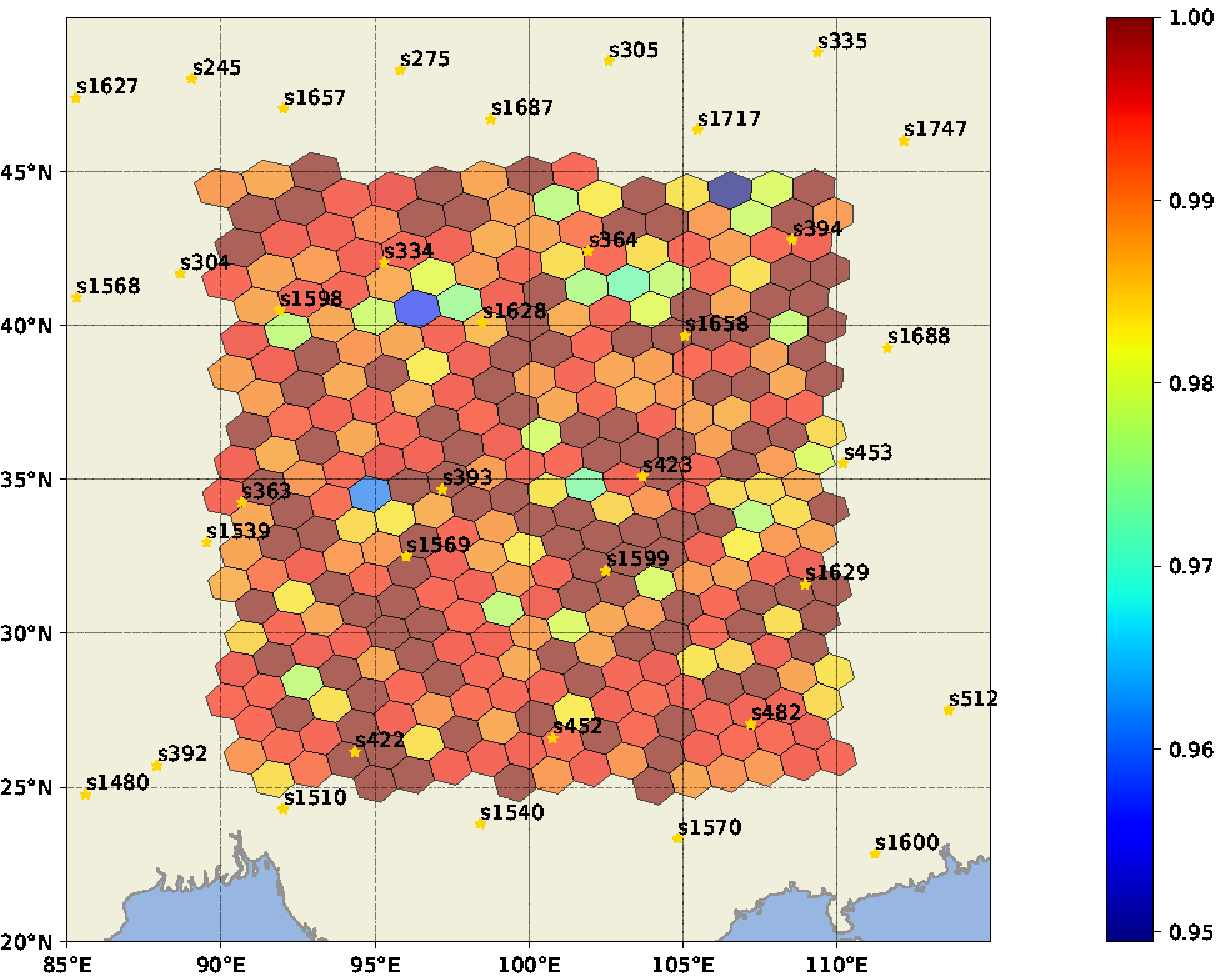} \vspace{-0.5em}
\caption{Access success probability for the Nearest scheme in target SCs.\label{fig:access_success_probability}} \vspace{-1.5em}
\end{figure}
The 5-th, 50-th, and 95-th percentile values for N-asset coverage, area traffic capacity, and service availability are listed in Table \ref{tab:constellation-KPI}. As can be seen, the 50\% N-asset coverage for SCs in the target areas is 9, which means a SC can have 9 service satellites whose signal quality is above a predefined threshold (i.e., SNR > -6dB in our simulation). Besides, the achieved area traffic capacity is around 4 Kbps/km$^2$ for the simulated target area. This value is quite smaller compared with the requirement (i.e., 10 Mbit/s/m$^2$ for hotspots) defined for terrestrial cellular networks.  The service availability ranges from 0.36 to 0.39, and thus each SC has nearly 1/3 time of satellite beam service. With respect to the unmet capacity performance, we randomly sample 10 SCs and plot the required data rate and offered data rate performance in Fig. \ref{fig:unmet_capacity}. It can be observed that the gap between offered and required data rate in some hotspot SCs, e.g., SC\#3, is quite large, and more than 20 percent of the total capacity demand is not satisfied.

The access related performance is plotted in Fig. \ref{fig:access_capacity} and Fig. \ref{fig:access_success_probability}. In particular, Fig. \ref{fig:access_capacity} shows the number of connected users per SC. As the number of users increases, the number of successfully connected users for both schemes grows. This indicates that higher collision rate due to more users is not dominant, and the larger available user pool finally leads to higher number of successful connected users. Besides, the SSB-plan-nearest scheme outperforms the Nearest method, since in the former case, users have the capability to smoothly switch to the next satellite when the current satellite is not available anymore. In contrast, for the latter case, users directly fallback to the idle state because no more satellites can be accessed in current time snapshot. Furthermore, Fig. \ref{fig:access_success_probability} plots the heat map in terms of the access success probability for the Nearest scheme. The success rates for all SCs exceed to $95\%$, which implies that almost all users can access the mega-constellation networks after few preamble retransmissions.

\begin{figure}[t]
\centering \vspace{-1.0em} \includegraphics[scale=0.45]{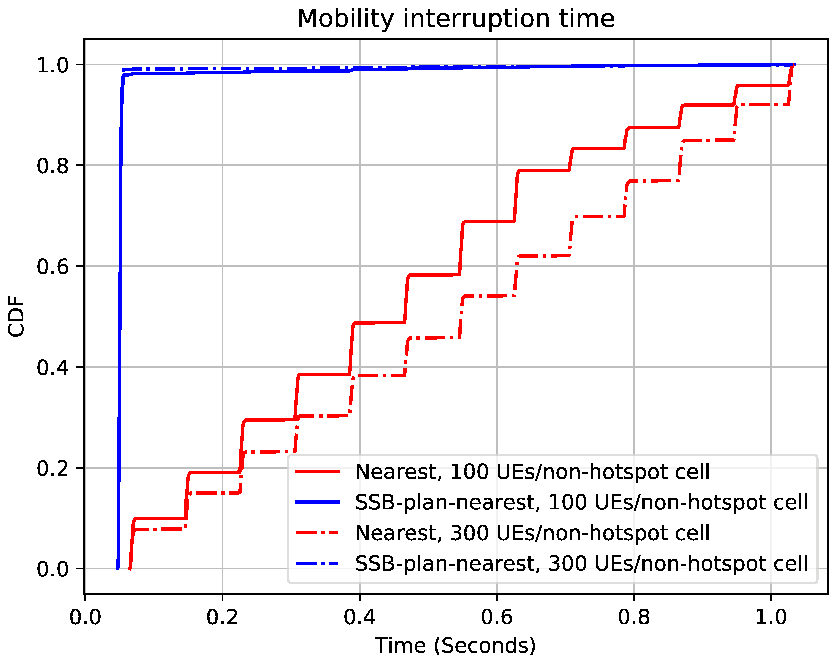} \vspace{-0.5em}
\caption{CDF of Mobility interruption time performance.\label{fig:Mobility_interruption_time} } \vspace{-1.0em}
\end{figure}

\begin{figure}[htp]
\textcolor{black}{}
\centering \vspace{-0.5em} \includegraphics[scale=0.32]{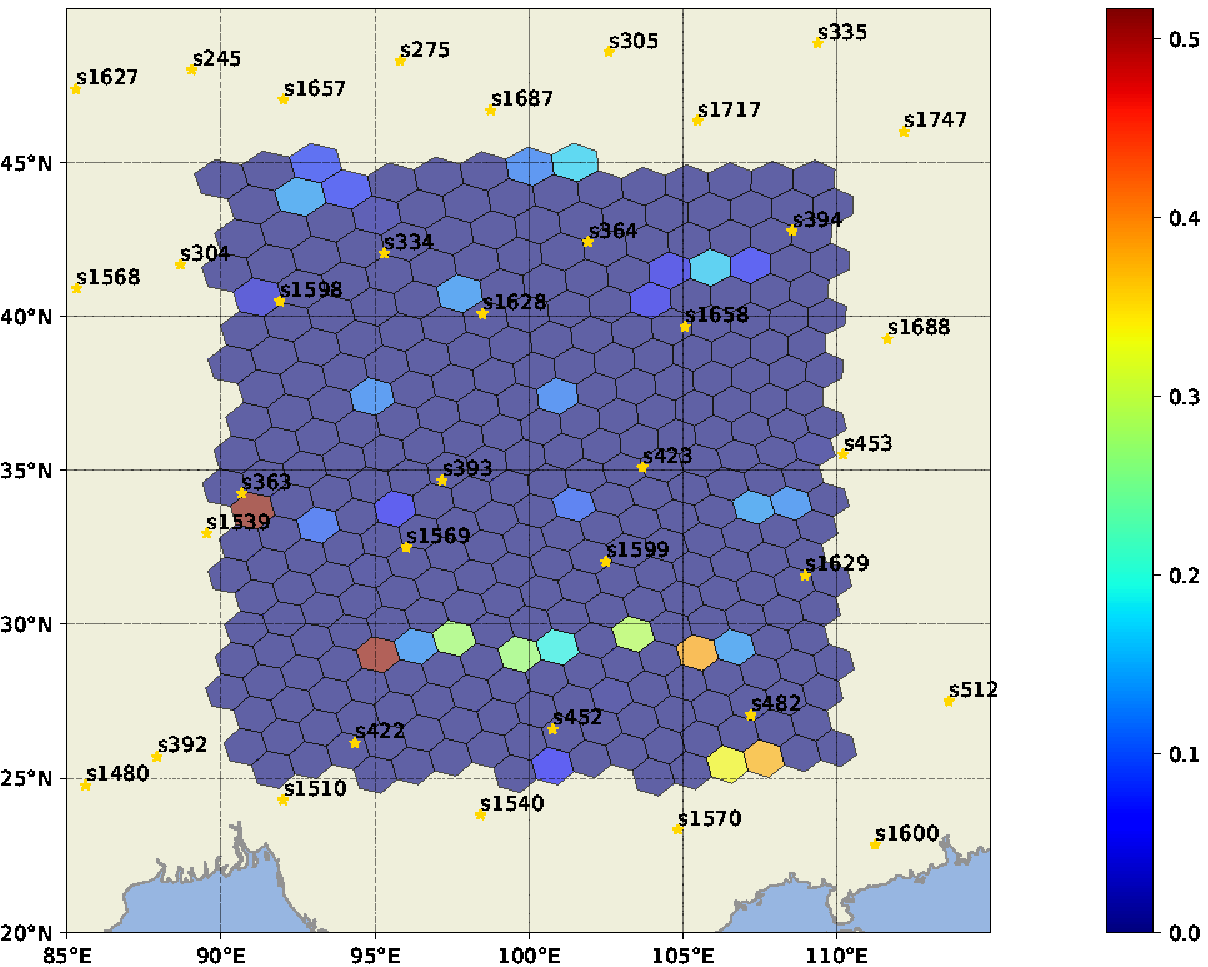}
\caption{HO failure rate for the SSB-plan-nearest scheme in target SCs.\label{fig:HO_failure_rate} } \vspace{-1.0em}
\end{figure}

The mobility related simulation results are illustrated in Fig. \ref{fig:Mobility_interruption_time} and Fig. \ref{fig:HO_failure_rate}. From Fig. \ref{fig:Mobility_interruption_time}, we observe that for the Nearest scheme, the mobility interruption time distributes uniformly between 0 and 1 second. This can be accounted by the fact that UEs can only take HOs to a target satellite within a time limit of 1s from the starting point of the upcoming snapshot. While in the SSB-plan-nearest scheme, most of UEs can initiate HO at current snapshot after measuring the SSB signal quality of a target satellite. In addition, a small portion of UEs make HOs from the starting point of the next snapshot when both the target satellites cannot provide service in the upcoming snapshot. Fig. \ref{fig:HO_failure_rate} depicts the heatmap of HO failure rate for all target SCs in the SSB-plan-nearest scheme. Notably, only a few SCs (especially the 10 hotspot cells) experience relatively higher HO failure rates (e.g., as much as $40\%$). This is due to that more UEs in the hotspots require HOs nearly at the same time. Therefore, the RO/preamble resources become a bottleneck during such kind of group HOs. \vspace{-0.5em}

\section{Conclusion\label{sec:Conclusion}}

In this paper, we have proposed a comprehensive KPI framework for LEO mega-constellation networks. An efficient multiple-satellite simulation scheme have been devised. Extensive system-level simulation results are present. We hope that the envisioned systematic KPI framework lays a solid foundation for further LMCSN-oriented performance assessment. As for future work, we tend to evaluate the KPI results for various LEO mega-constellation configurations, different types of traffic data, and varying duration of time snapshot. Besides, potential LMCSN specific technologies can be investigated to improve the network performance, e.g., area traffic capacity and HO failure rate.

\bibliographystyle{IEEEtran}
\bibliography{ref_NTN_KPI}

\begin{thebibliography}{10}
\providecommand{\url}[1]{#1}
\csname url@samestyle\endcsname
\providecommand{\newblock}{\relax}
\providecommand{\bibinfo}[2]{#2}
\providecommand{\BIBentrySTDinterwordspacing}{\spaceskip=0pt\relax}
\providecommand{\BIBentryALTinterwordstretchfactor}{4}
\providecommand{\BIBentryALTinterwordspacing}{\spaceskip=\fontdimen2\font plus
\BIBentryALTinterwordstretchfactor\fontdimen3\font minus
  \fontdimen4\font\relax}
\providecommand{\BIBforeignlanguage}[2]{{%
\expandafter\ifx\csname l@#1\endcsname\relax
\typeout{** WARNING: IEEEtran.bst: No hyphenation pattern has been}%
\typeout{** loaded for the language `#1'. Using the pattern for}%
\typeout{** the default language instead.}%
\else
\language=\csname l@#1\endcsname
\fi
#2}}
\providecommand{\BIBdecl}{\relax}
\BIBdecl

\bibitem{Survey_NTN5Gto6G}
M.~M. Azari and \emph{et al.}, ``Evolution of non-terrestrial networks from
  5{G} to 6{G}: A survey,'' \emph{IEEE Communications Surveys \& Tutorials},
  vol.~24, no.~4, pp. 2633--2672, Fourth Quarter, 2022.

\bibitem{Survey_SatComintheNewSpaceEra}
O.~Kodheli and \emph{et al.}, ``Satellite communications in the new space era:
  A survey and future challenges,'' \emph{IEEE Communications Surveys \&
  Tutorials}, vol.~23, no.~1, pp. 70--109, First Quarter, 2021.

\bibitem{VLEO_HW}
H.~Luo and \emph{et al.}, ``{6G VLEO} satellite networks,''
  \emph{Communications of Huawei Research}, vol.~2, pp. 34--45, Sep. 2022.

\bibitem{38821_SolutionstoSupportNTN}
{3GPP TR 38.821 V16.0.0}, ``Solutions for {NR} to support non-terrestrial
  networks ({NTN}),'' Release 16, Dec. 2019.

\bibitem{38811}
{3GPP TR 38.811 V15.0.0}, ``Study on new radio ({NR}) to support non
  terrestrial networks,'' Release 15, Jun. 2018.

\bibitem{ITUR_M2514}
{Report ITU-R M.2514-0}, ``Vision, requirements and evaluation guidelines for
  satellite radio interface(s) of imt-2020,'' Sep. 2022.

\bibitem{38913}
{3GPP TR 38.913 V17.0.0}, ``Study on scenarios and requirements for next
  generation access technologies,'' Release 17, Mar. 2022.

\bibitem{Nokia_HOSolutionsforLEO}
E.~Juan, M.~Lauridsen, J.~Wigard, and P.~Mogensen, ``Handover solutions for
  5{G} low-earth orbit satellite networks,'' \emph{IEEE Access}, vol.~10, pp.
  2169--3536, Aug. 2022.

\bibitem{ConstellationDesign}
I.~Leyva-Mayorga and \emph{et al.}, ``{NGSO} constellation design for global
  connectivity,'' \emph{arXiv:2203.16597}, pp. 1--26, Apr. 2022.

\bibitem{ChinaCom_6GServiceCoverage}
M.~Sheng and \emph{et al.}, ``6{G} service coverage with mega satellite
  constellations,'' \emph{China Commu.}, vol.~19, no.~1, pp. 64--76, Jan. 2022.

\bibitem{WCM_SAGIN_Simulation}
N.~Chen and \emph{et al.}, ``A comprehensive simulation platform for
  space-air-ground integrated network,'' \emph{IEEE Wireless Commu.}, vol.~27,
  no.~1, pp. 178--185, Feb. 2020.

\bibitem{ICC_CooperativeBeamHopping}
Y.~Wang and \emph{et al.}, ``Cooperative beam hopping for accurate positioning
  in ultra-dense leo satellite networks,'' in \emph{IEEE ICC Workshops}, Jul.
  2022.

\bibitem{BH_Simulation}
J.~Zhang and \emph{et al.}, ``System-level evaluation of beam hopping in
  {NR}-based {LEO} satellite communication system,'' in \emph{IEEE WCNC}, Mar.
  2023.

\bibitem{H3}
https://h3geo.org.

\bibitem{TS38331}
{3GPP TS 38.331 V17.2.0}, ``{NR} radio resource control ({RRC}) protocol
  specification,'' Release 17, Sep. 2022.

\end{thebibliography}

\end{document}